\pdfoutput=1
\documentclass[reprint,aip,jcp,twocolumn,showpacs]{revtex4-1}
\usepackage[paperwidth=210mm,paperheight=297mm,centering,hmargin=2.0cm,vmargin=2.5cm]{geometry}

\newcommand{\eref}[1]{Eq.~(\ref{#1})}%
\newcommand{\Eref}[1]{Equation~(\ref{#1})}%
\newcommand{\fref}[1]{Fig.~\ref{#1}} %

\usepackage{natbib}
\usepackage{graphicx}
\usepackage[utf8]{inputenc}
\usepackage{mathptmx}
\usepackage{mathtools}
\usepackage{epsfig,amsopn}
\usepackage{graphicx}
\usepackage{amsmath,amssymb}
\usepackage{xcolor}
\usepackage{braket}
\usepackage{float}
\usepackage{enumerate}
\usepackage[symbol]{footmisc}
\usepackage{appendix}
\usepackage{stackengine}
\usepackage[normalem]{ulem}
\usepackage{todonotes}
\allowdisplaybreaks[1]

\def\bea{\begin{eqnarray}}
\def\eea{\end{eqnarray}}

\begin{document}
%%%%%%%%%%%%%%%%%%%%%%%  Normal format %%%%%%%%%%%%%%%%%%%%%%%%%%%%%%%%%
\title{Resetting transition is governed by an interplay between thermal and potential energy}

\author{Somrita Ray}
\email{somrita@iittp.ac.in}
\affiliation{\noindent \textit{School of Chemistry, The Center for Physics and Chemistry of Living Systems, The Raymond and Beverly Sackler Center for Computational Molecular and Materials Science, \& The Ratner Center for Single Molecule Science, Tel Aviv University, Tel Aviv 69978, Israel}}
\affiliation{\noindent \textit{Department of Chemistry, Indian Institute of Technology Tirupati, Tirupati 517506, India}}
\author{Shlomi Reuveni}
\email{shlomire@tauex.tau.ac.il}
\affiliation{\noindent \textit{School of Chemistry, The Center for Physics and Chemistry of Living Systems, The Raymond and Beverly Sackler Center for Computational Molecular and Materials Science, \& The Ratner Center for Single Molecule Science, Tel Aviv University, Tel Aviv 69978, Israel}}

\date{\today}
%%%%%%%%%%%%%%%%%%%%%%%%%%%%%%%%%%%%%%%%%%%%%%%%%%%%%%%%%%%%%%%%%%%%%%%

\begin{abstract}
\noindent
A dynamical process that takes a random time to complete, e.g., a chemical reaction, may  either be accelerated or hindered due to resetting. Tuning system parameters such as temperature, viscosity or concentration, can invert the effect of resetting on the mean completion time of the process, which leads to a resetting transition. Though the resetting transition was recently studied for diffusion in a handful of model potentials, it is yet unknown whether the results follow any universality in terms of well-defined physical parameters. To bridge this gap, we propose a general framework which reveals that the resetting transition is governed by an interplay between thermal and potential energy. This result is illustrated for different classes of potentials that are used to model a wide variety of stochastic processes with numerous applications.
\end{abstract}

%\keywords{Stochastic resetting, resetting transition}

\maketitle

\begin{figure}[t!]
\begin{center}
\includegraphics[width=8.25cm]{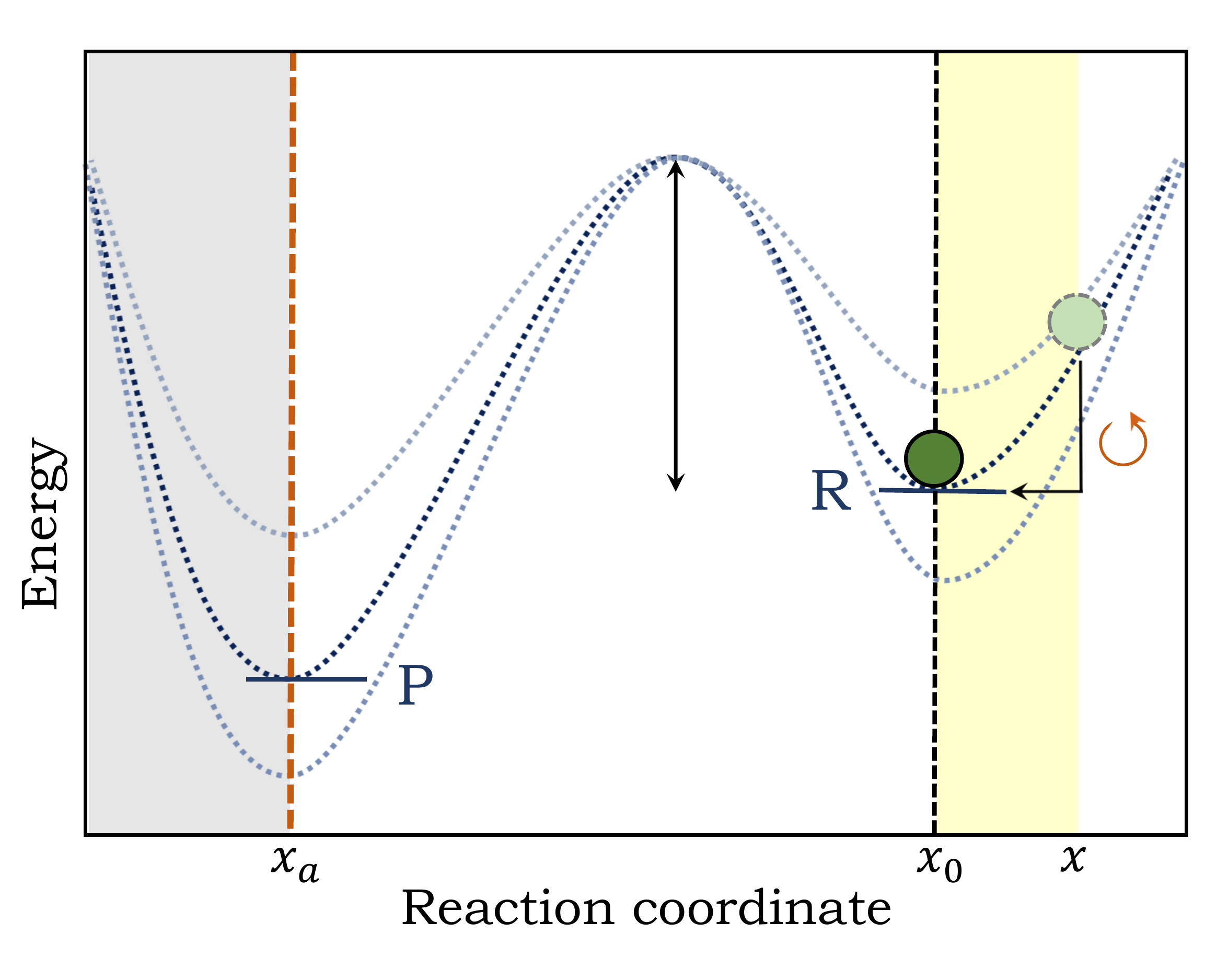}
\end{center}
\caption{Schematic illustration of a chemical reaction $R\to P$ envisioned as diffusion of a Brownian particle in a double well potential. The completion time of the reaction is given by the particle's  first-passage time from $x_0$ (state $R$) to $x_a$ (state $P$). Stochastic resetting of the reaction brings the reaction coordinate back to $x_0$ at random time epochs. Depending on parameters, resetting can either increase or decrease the mean time taken to form a product.}
\label{Model}
\end{figure}

%%%%%%%%%%%%%%%%%%%%%%%%%%%%%%%%%%%%%%%%%%%%%%%%%%%
Resetting refers to a situation where an ongoing dynamical process is stopped in its midst and started over \cite{SR1,SR2,SR3,ReuveniPRL,PalReuveniPRL,HomeRangeSearch1,ReviewSNM}. Theoretical study of stochastic dynamics with resetting has drawn overwhelming attention in recent years\cite{SP1,SP2,SP3,SP4,SP5,SP6,SP7,SP8,SP9,SP10,SP11,SP12,SP13,SP14,SP15,SP16,SP17,SP18,SP19,SP20,SP21,SP22,SP23,SP24,SP25,SP26,SP27,SP28,SP29,SP30} and applications to problems in chemical \cite{RB1,RB2,RB3,RB4,RB5} and biological \cite{BP1,BP2,BP3,BP4,BP5,BP6,BP7} physics have further amplified activity in this rapidly emerging field. In addition, experimental realizations of diffusion with resetting have been successful very recently, where optical tweezers \cite{RExp1} or laser traps \cite{RExp2} were used to reset the position of a colloidal particle. These studies opened up further possibilities to explore the dynamics of physical systems with stochastic resetting.\\
\indent
To illustrate resetting, consider a chemical reaction $R\to P$ that owns an energy profile with two minima, where the reactant ($R$) and product ($P$) states are located. This system can be modeled as diffusion in a double-well potential, where the reaction coordinate is mapped onto the position of a diffusing particle (\fref{Model}). The time to complete the reaction is then given by the random time it takes the particle to get from one minimum of the potential to the other, across a separating energy barrier. This time is commonly known as the {\it first-passage time} \cite{gardiner,cox,redner,persistence,FPReview,zilman}. 

%%%%%---MODEL FIGURE----%%%%%%%%%%%%
%%%%%%%%%%%%%%%%%%%%%
\indent
Chemical reactions, such as the one described in \fref{Model}, are naturally subject to stochastic resetting. Resetting happens e.g., when two species that bind to form an activated complex, unbind without actually reacting; but only to rebind again at some later time. In a similar way, unbinding acts to reset enzymatic reactions and a one-on-one mapping between first-passage with resetting and the Michaelis-Menten reaction scheme has recently been established \cite{RB1,RB2,RB3,RB4}. Chemical reactions can also be reset by means of external manipulation, e.g., an electromagnetic pulse can bring the reactant to its initial state [($R$) in \fref{Model}]. Importantly, it has been found that resetting can either hinder or expedite the completion of a chemical reaction (or any other first-passage %time
 process), with the net effect determined by various physical parameters such as temperature, viscosity, and the potential energy landscape underlying the reaction at hand. Tuning one (or more) of these governing parameters across some critical value can invert the effect of resetting on the mean completion time of the reaction, which leads to a ``resetting transition''\cite{RB1,RB2,exponent,Landau}.\\
\indent
In recent years, the effect of resetting on diffusion in various model potentials, e.g., linear, harmonic, power-law and logarithmic \cite{RayReuveniJPhysA,exponent,logpotential,SDDR} was thoroughly explored. In each of these studies, however, the resetting transition was characterized in terms of a different physical governing parameter that was specific to the system under consideration. It is thus still unclear whether one can present a unifying physical understanding of the resetting transition that applies to diffusion in arbitrary potentials. Motivated by this question, we develop a general approach to show that the resetting transition can be understood in terms of an interplay between two competing energy scales, viz., the  potential energy and the thermal energy. Our results are general and we illustrate them with three different classes of potentials that are used to model a wide variety of stochastic processes with numerous applications. \\
\indent
We start with a first-passage process, e.g., the chemical reaction shown in Fig.1, but this time we consider an arbitrary energy profile $U(x)$ to keep things general. This scenario can be modeled with a particle undergoing diffusion with a diffusion coefficient $D$ in an arbitrary potential $U(x)$. Letting $p(x,t|x_0)$ denote the conditional probability density of finding the particle at a position $x$ at time $t$, provided its initial position was $x_0>0$, we write the backward Fokker Planck equation for the process \cite{gardiner,cox,redner}
\begin{eqnarray}
\dfrac{\partial p(x,t|x_0)}{\partial t}=
\left[\frac{-U^{\prime}(x_0)}{\zeta}\right]\dfrac{\partial p(x,t|x_0)}{\partial x_0}+D\dfrac{\partial^2 p(x,t|x_0)}{\partial x_0^{2}},\nonumber\\
\label{eq:bfpe}
\end{eqnarray}
\noindent
where $\zeta$ denotes the friction coefficient and $U^{\prime}(x)\equiv dU(x)/dx$.
The process ends when the particle hits an absorbing boundary, which is placed on the non-negative axis to the right side of the origin at $0\le x_a\le x_0$. When this happens, the particle is removed from the system which leads to the boundary condition $p(x_a,t|x_0) = 0$. The survival probability, i.e., the probability of finding the particle within the interval $[x_a,\infty)$ %(considering $x_0\geq x_a$) 
at time $t$, is given by $Q(t|x_0)=\int_{x_{a}}^{\infty}p(x,t|x_0)dx$. Integrating \eref{eq:bfpe} with respect to $x$ over $[x_a,\infty)$ thus gives
\begin{eqnarray}
\dfrac{\partial Q(t|x_0)}{\partial t}=
\left[\frac{-U^{\prime}(x_0)}{\zeta}\right]\dfrac{\partial Q(t|x_0)}{\partial x_0}+D\dfrac{\partial^2 Q(t|x_0)}{\partial x_0^{2}}.%\nonumber\\
\label{eq:spde}
\end{eqnarray}
\noindent
Here the initial condition is $Q(0|x_0)=1$ and the boundary condition is $Q(t|x_a) = 0$. Letting $T$ denote the first-passage time (FPT) to the absorbing boundary placed at $x_a$, we recall that the probability density function of this random variable\cite{gardiner,cox,redner} is given by $-\partial Q(t|x_0)/\partial t$. Thus the $n$-th moment of the FPT is $\tau_n=\left<T^n\right>\coloneqq-\int_{0}^{\infty}dt~t^{n}\left[\frac{\partial~Q(t|x_0)}{\partial~t}\right]=n\int_{0}^{\infty}dt~t^{n-1} Q(t|x_0)$, where the second equality comes from integrating by parts and further demanding that $\left.Q(t|x_0)\right|_{t\to\infty}=0$. Multiplying \eref{eq:spde} by $t^{n-1}$ and integrating over $[0,\infty]$ with respect to $t$ we get\cite{gardiner,cox}
\begin{align}
-n\tau_{n-1}=-\left[\frac{U^{\prime}(x_0)}{\zeta}\right]\dfrac{d \tau_n}{d x_0}+D\dfrac{d^2 \tau_n}{d x_0^{2}},
\label{eq:tnde}
\end{align}
where we utilize the previous identities to identify the moments.\\
\indent
\eref{eq:tnde} is a single-variable recursive differential equation in $\tau_n$. Given $\tau_{n-1}$, it can be solved with appropriate boundary conditions \cite{gardiner,cox}. This method allows us to calculate the moments of the FPT directly, bypassing the calculation of the survival probability $Q(t|x_0)$. Taking into account the fact that by definition $\tau_0=1$, we solve \eref{eq:tnde} for $n=1$ to obtain an expression for the mean FPT \cite{gardiner} $\tau_1=D^{-1}\int_{x_a}^{x_0}dy~e^{\beta U(y)}\int_{y}^{\infty}dz~e^{-\beta U(z)}$ [see the Supporting Information for the derivation],
where we utilized the Einstein relation, $D=(\beta \zeta)^{-1}$. In a similar spirit, setting $n=2$ in \eref{eq:tnde} and plugging in the expression of $\tau_1$ into it, we get a differential equation in $\tau_2$, whose solution reads 
$\tau_2=2D^{-2}\int_{x_a}^{x_0}dw~e^{\beta U(w)}
\int_{w}^{\infty}dv~e^{-\beta U(v)}\int_{x_a}^{v}dy~e^{\beta U(y)}\int_{y}^{\infty}dz~e^{-\beta U(z)}$ [see the Supporting Information].
%\int_{x_a}^{v}dy~e^{\beta U(y)}\int_{y}^{\infty}dz~e^{-\beta U(z)}]/D^2$
In what follows, we will discuss how $\tau_1$ and $\tau_2$ together can predict the effect of resetting, and relate the latter to physical governing parameters.\\
%%%%%%%%%%%%%%%%%%%%%%%%%%%%%%%%%%%%%%%%%%%%%%%%%%%%%%%%%%%%%%%%%%%%%%%%%%%%%%%%%%%%%%%%%%%%%%%%%%%%%%%
%\indent\underline{\it Condition for resetting transition:}\hspace{0.4cm}
%%%%%%%%%%%%%%%%%%%%%%%%%%%%%%%%%%%%%%%%%%%%%%%%%%%%%%
\indent
The theory of first-passage with resetting allows us to express the FPT of a process with resetting in terms of the FPT of the same process without resetting\cite{RB1,ReuveniPRL,PalReuveniPRL}. The effect of resetting can thus be predicted based on the FPT of the underlying process. In particular, it was shown that the introduction of stochastic resetting accelerates first-passage whenever the standard deviation of the FPT is greater than its mean, i.e, $\sqrt{\tau_2-\tau_1^2}>\tau_1$. In stark contrast, first-passage is delayed due to the introduction of stochastic resetting whenever $\sqrt{\tau_2-\tau_1^2}<\tau_1$. In real-life systems, the FPT distribution and hence its moments $\tau_1$ and $\tau_2$ will vary when the governing parameters are altered. This indicates that by tuning physical parameters like temperature, the effect of resetting on the FPT can be inverted. The associated transition, i.e., the resetting transition \cite{exponent,Landau,RayReuveniJPhysA,logpotential,SDDR} occurs when $\sqrt{\tau_2-\tau_1^2}=\tau_1$, which is equivalent to $\tau_2=2\tau_1^2$. Note that if either the mean or the standard deviation of the FPT diverges, the introduction of stochastic resetting always accelerates first-passage. Thus, we exclude these cases from the discussion below and focus only on situations where the mean and second moment of the FPT are finite. \\
\indent
Utilizing the expressions for $\tau_1$ and $\tau_2$ above, we see that the resetting transition of diffusion in a potential landscape $U(x)$ occurs when
\begin{widetext} %%%%%%%%%%%%%%%%%%%%%%%%%%%%%%%% WIDE TEXT BEGINS %%%%%%%%%%%%%%%%%%%%%%%%%%%%
\begin{eqnarray}
\int_{x_a}^{x_0}dw~e^{+\beta U(w)}\int_{w}^{\infty}dv~e^{-\beta U(v)}\int_{x_a}^{v}dy~e^{+\beta U(y)}\int_{y}^{\infty}dz~e^{-\beta U(z)}=
\left[\int_{x_a}^{x_0}dy~e^{+\beta U(y)}\int_{y}^{\infty}dz~e^{-\beta U(z)}\right]^2.
\label{eq:res_trans}
\end{eqnarray}
 %%%%%%%%%%%%%%%%%%%---LP----%%%%%%%%%%%%
\begin{figure*}[t!]
\begin{center}
\includegraphics[width=5.5cm]{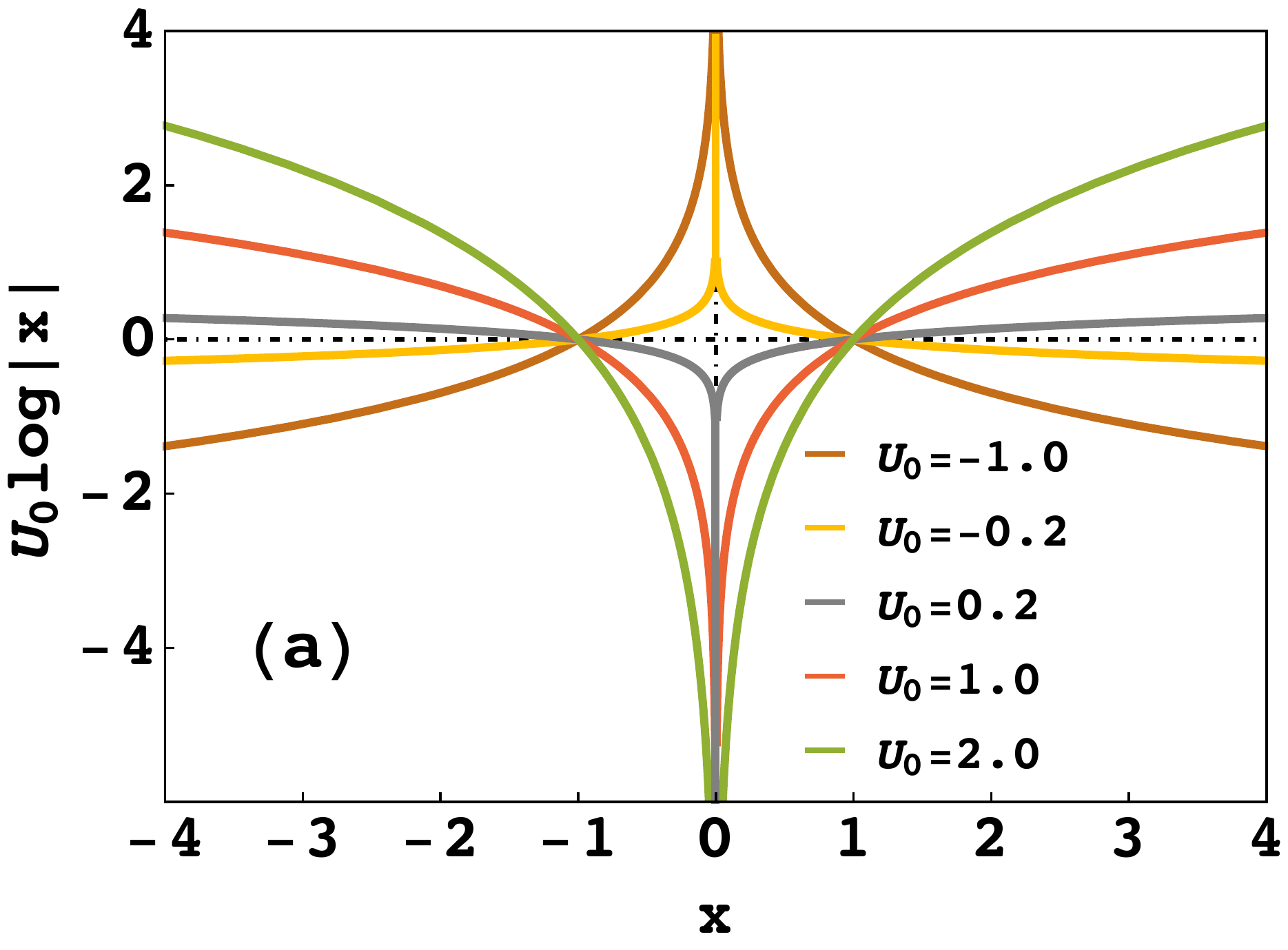}
\includegraphics[width=5.66cm]{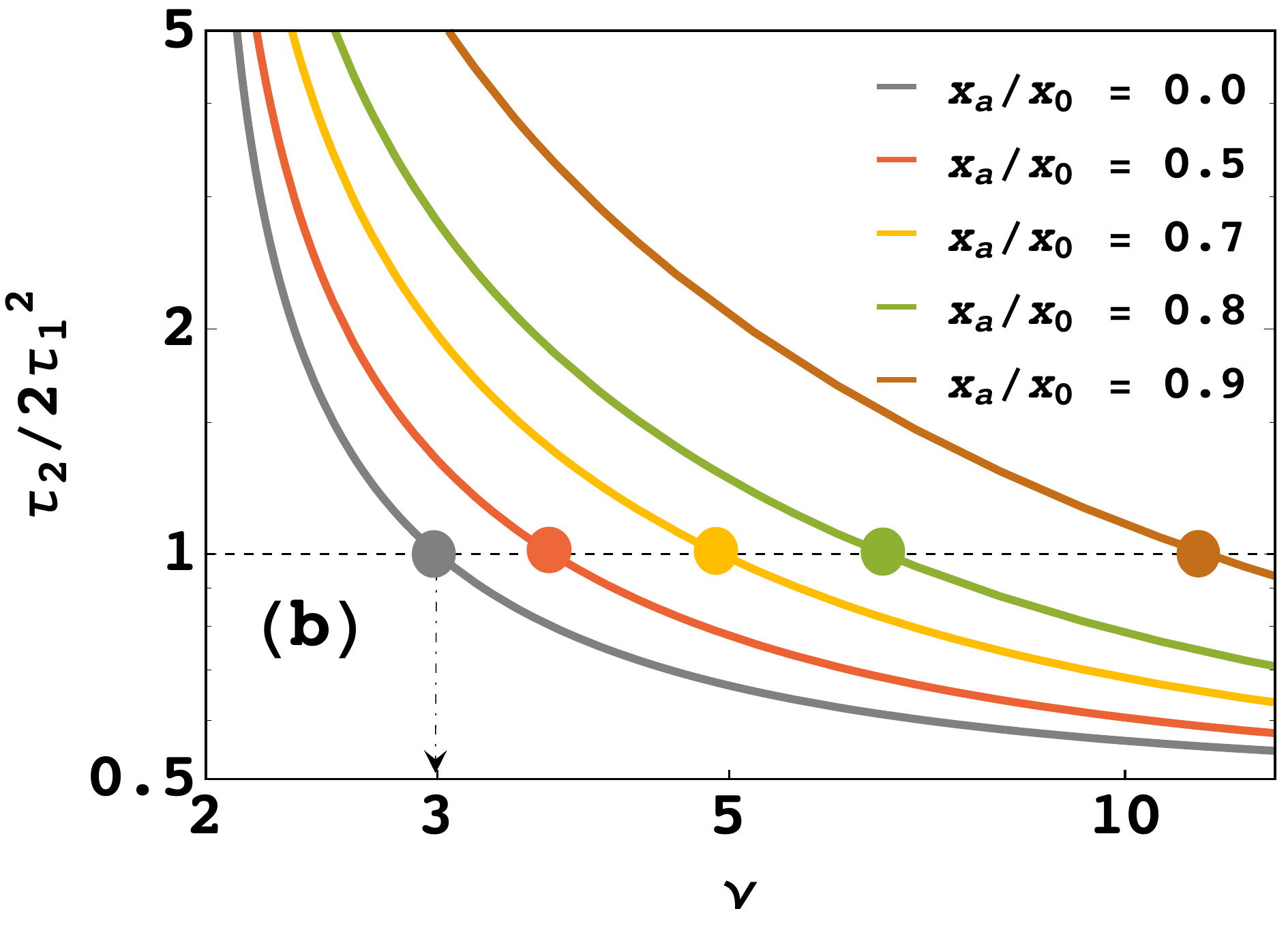}
\includegraphics[width=5.48cm]{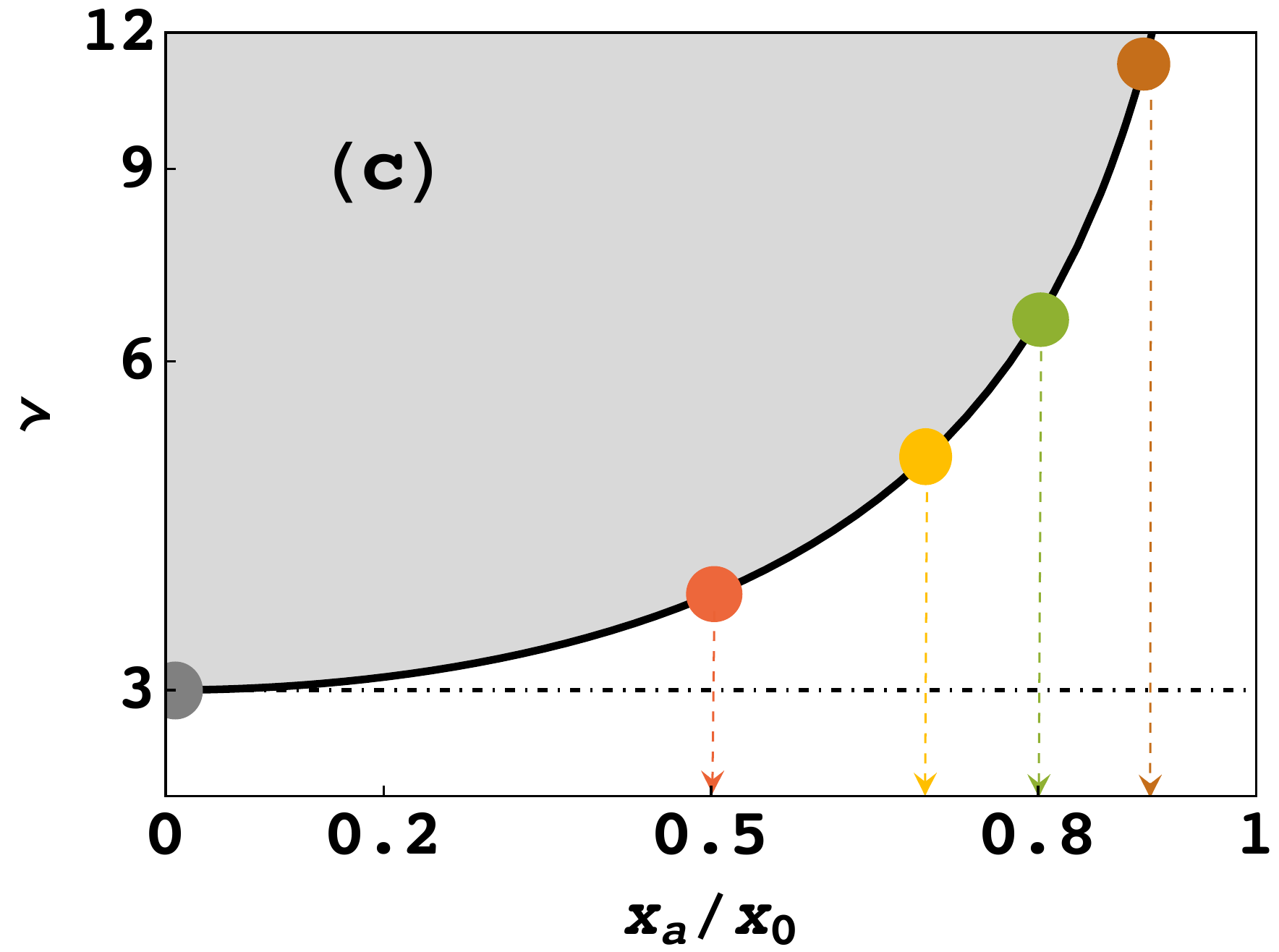}
\end{center}
\caption{Panel (a): A logarithmic potential $U(x)=U_0\log|x|$ for different values of $U_0$. $U(x)$ is repulsive for $U_0<0$ and attractive for $U_0>0$. The potential has a central singularity at $x=0$ and it can be effectively regularized by placing an absorbing boundary at $x_a>0$. Panel (b): Graphical solution of \eref{eq:cv_log} for different values of $x_a/x_0$. The solutions, denoted $\nu^{\star}$, are the abscissa corresponding to the colored circles. Panel (c): Phase diagram showing the effect of resetting on first-passage.  The white region marks the part of the phase space where the introduction of resetting expedites first-passage, while the shaded region marks the opposite. The black line, obtained by plotting $\nu^{\star}$ vs. $x_a/x_0$ using \eref{eq:nu_star}, separates the two phases and indicates the points of the resetting transition for any given value of $x_a/x_0$. Circles of the same colors show same values of $x_a/x_0$ as in panel (b). }
\label{FigLP}
\end{figure*}
\end{widetext}
%%%%%%%%%%%%%%%%%%%%% WIDE TEXT ENDS%%%%%%%%%%%%%%%%%%%
\noindent
We will now show that \eref{eq:res_trans} can be used to characterize the resetting transition in terms of an interplay between the thermal energy $\beta^{-1}$ and the potential energy. We do this by considering diffusion in various canonical potentials, starting with the logarithmic potential.\\
%%%%%%%%%%%%%%%%%%%%%%%%%%%%%%%%%%%%%%%%%%%
%\indent \underline{\it Logarithmic potential:}\hspace{0.1cm} 
%%%%%%%%%%%%%%%%%%%%%%%%%%%%%%%%%%%%%%%%%%%%%%%%%%%%%%%%%%%%%%%%%%%%%%%%%%%%%%%%%%%%%%%
\indent 
Diffusion in a logarithmic potential is a popular choice to model stochastic phenomena such as the denaturation of double-stranded DNA by bubble formation\cite{denaturation1,denaturation2}, interactions of colloids and polymers with walls of narrow channels and pores\cite{entropicpotential1,entropicpotential2,entropicpotential3,entropicpotential4}, and spreading of momenta of cold atoms trapped in optical lattices\cite{opticallattice1,opticallattice2,opticallattice3}. A simple log potential of the form $U(x)=U_0\log|x|$ (where $U_0$ is the strength of the potential) owns a singularity at $x=0$ [\fref{FigLP}(a)]. In many physical applications, however, one frequently encounters versions of the logarithmic potential sans the central singularity, commonly known as ``regularized'' log potentials. 
One simple trick to regularize the potential $U(x)=U_0\log|x|$ is to place the absorbing boundary at a position $x_a>0$.  \\
\indent
In a recent study\cite{logpotential}, we comprehensively analysed the effect of resetting on diffusion in a log potential (commonly known as the Bessel process \cite{AJBray,besselFPT}) for the special case of $x_a=0$, i.e, where the absorbing boundary is placed at the origin. We demonstrated that the system displays different dynamical behaviors as the dimensionless parameter $\beta U_0$, the strength of the potential in units of the thermal energy, is tuned. In particular, we proved that resetting expedites the first-passage to the origin when $\beta U_0<5$, but delays the same when $\beta U_0>5$, leading to a resetting transition at $\beta U_0=5$. The present framework allows us to generalize this result to a regularized log potential, for which $0<x_a<x_0$.\\
\indent
Setting $U(x)=U_0\log|x|$ in \eref{eq:res_trans}, we get $e^{\pm\beta U(x)}=|x|^{\pm\beta U_0}$, which allows us to analytically evaluate the integrals. Doing so, we obtain
\begin{align}
\frac{\tau_2}{2\tau_1^2}=\frac{1}{2}\left[1-\frac{x_a^2+x_0^2}{(x_a^2-x_0^2)(\nu-2)}\right]
=1,
\label{eq:cv_log}
\end{align}
\noindent
where $\nu\coloneqq(1+\beta U_0)/2$ is the system parameter that governs the resetting transition by capturing the interplay of the energy scales $\beta^{-1}$ and $U_0$. In \fref{FigLP}(b), we graphically solve \eref{eq:cv_log} by plotting its left hand side vs. $\nu$. The solutions, denoted $\nu^{\star}$, are functions of the ratio $x_a/x_0$ and are given by
\begin{eqnarray}
\nu^{\star} = 2+ \left[\frac{1+(x_a/x_0)^2}{1-(x_a/x_0)^2}\right].
\label{eq:nu_star}
\end{eqnarray}
Thus $\nu^{\star}$ is the critical value of $\nu$ at which the resetting transition occurs. \Eref{eq:nu_star} clearly indicates that in the limit $x_a\to 0$, the resetting transition occurs at $\nu^{\star}=3$ (i.e., $\beta U_0 =5$), which agrees with our earlier result \cite{logpotential}.\\
\indent
In \fref{FigLP}(c), we plot $\nu^{\star}$ as a function of $x_a/x_0$ (black line), which separates the phase space in two regions with respect to the effect of resetting. Recalling the definition of $\nu$, we see that $\nu$ is large for large values of $\beta U_0>0$. The strongly attractive potential then drives the diffusing particle towards the origin, i.e, in the direction of the absorbing boundary. Resetting to $x_0$ thus delays its first-passage by interrupting that driven motion (shaded region in \fref{FigLP}(c)). In contrast, $\nu$ is small for small values of $\beta U_0>0$, i.e, when the drive towards the origin is weak. In that case, thermal diffusion predominates and introduction of resetting expedites the particle's first-passage by effectively reducing the possibilities of it diffusing away from the origin (white region in \fref{FigLP}(c)). The black line given by \eref{eq:nu_star} characterizes the condition for resetting transition by identifying the precise separatrix between the above mentioned phases, one where potential energy dominates and the other where thermal energy dominates. In the present example, we see that the separatrix depends on the position of the absorbing boundary relative to the particle's initial position.  \\
%%%%%%%---PLP----%%%%%%%%%%%%%%%
\begin{figure*}[t!]
\begin{center}
\includegraphics[width=5.5cm]{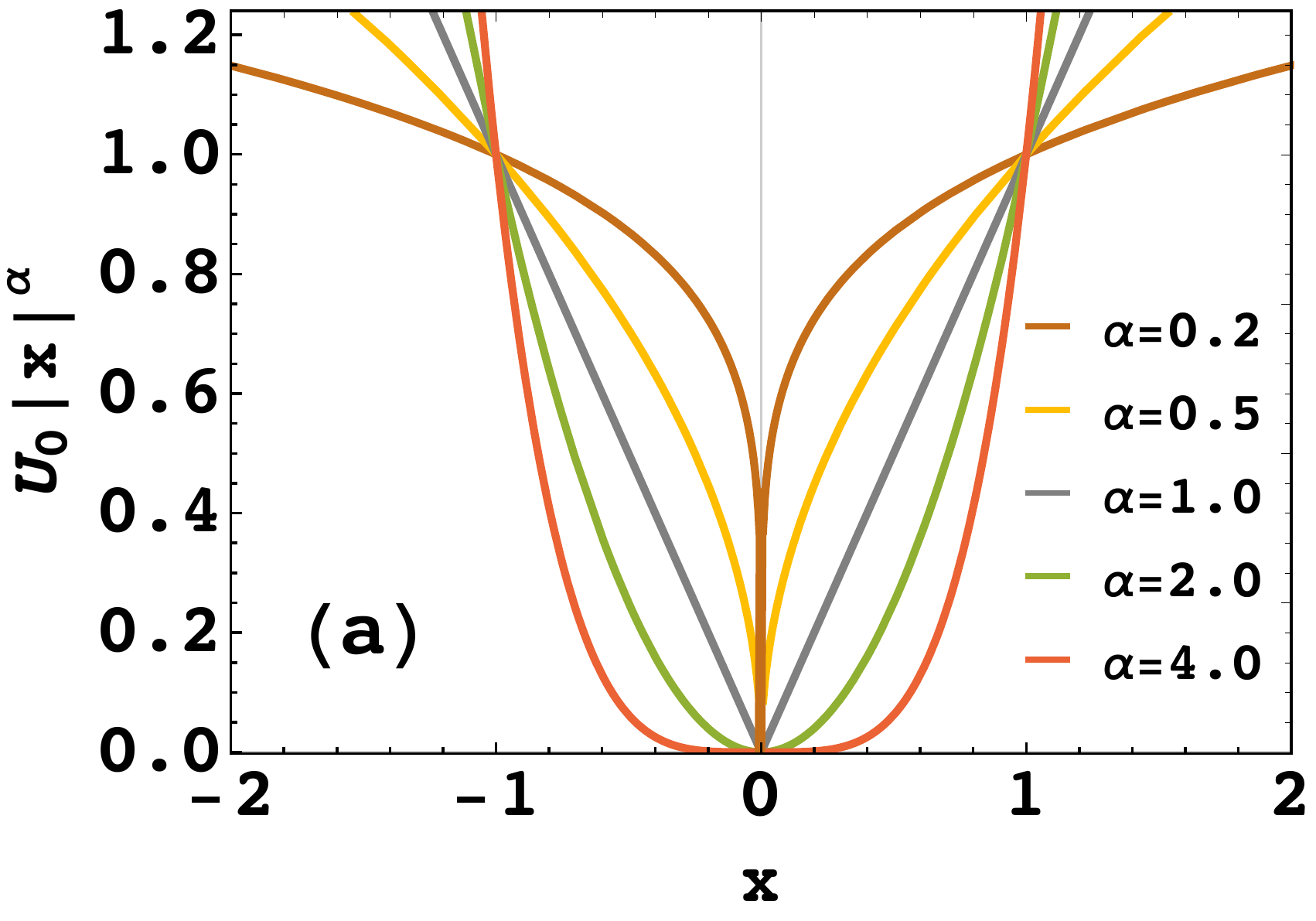}
\includegraphics[width=5.4cm]{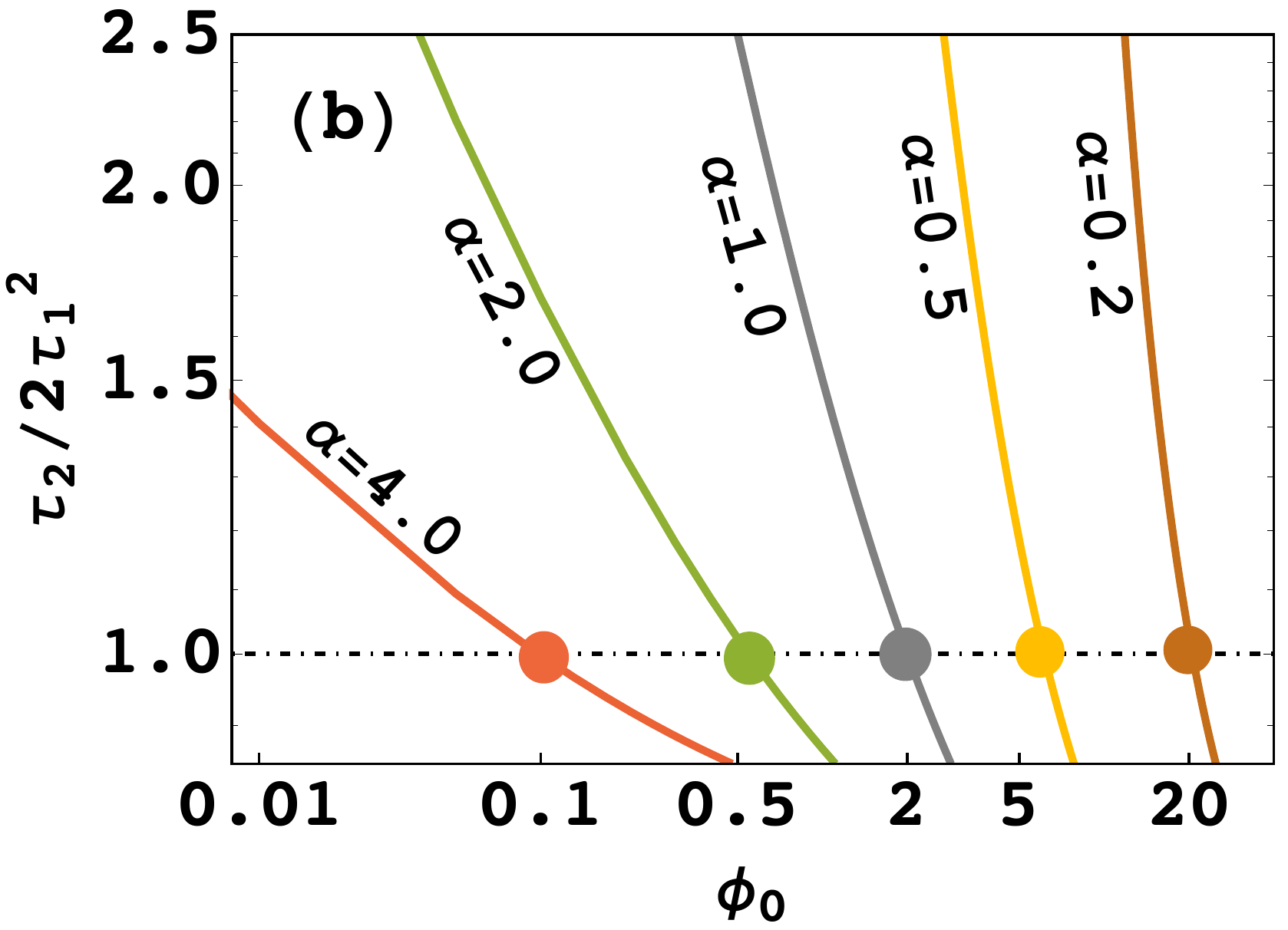}
\includegraphics[width=5.3cm]{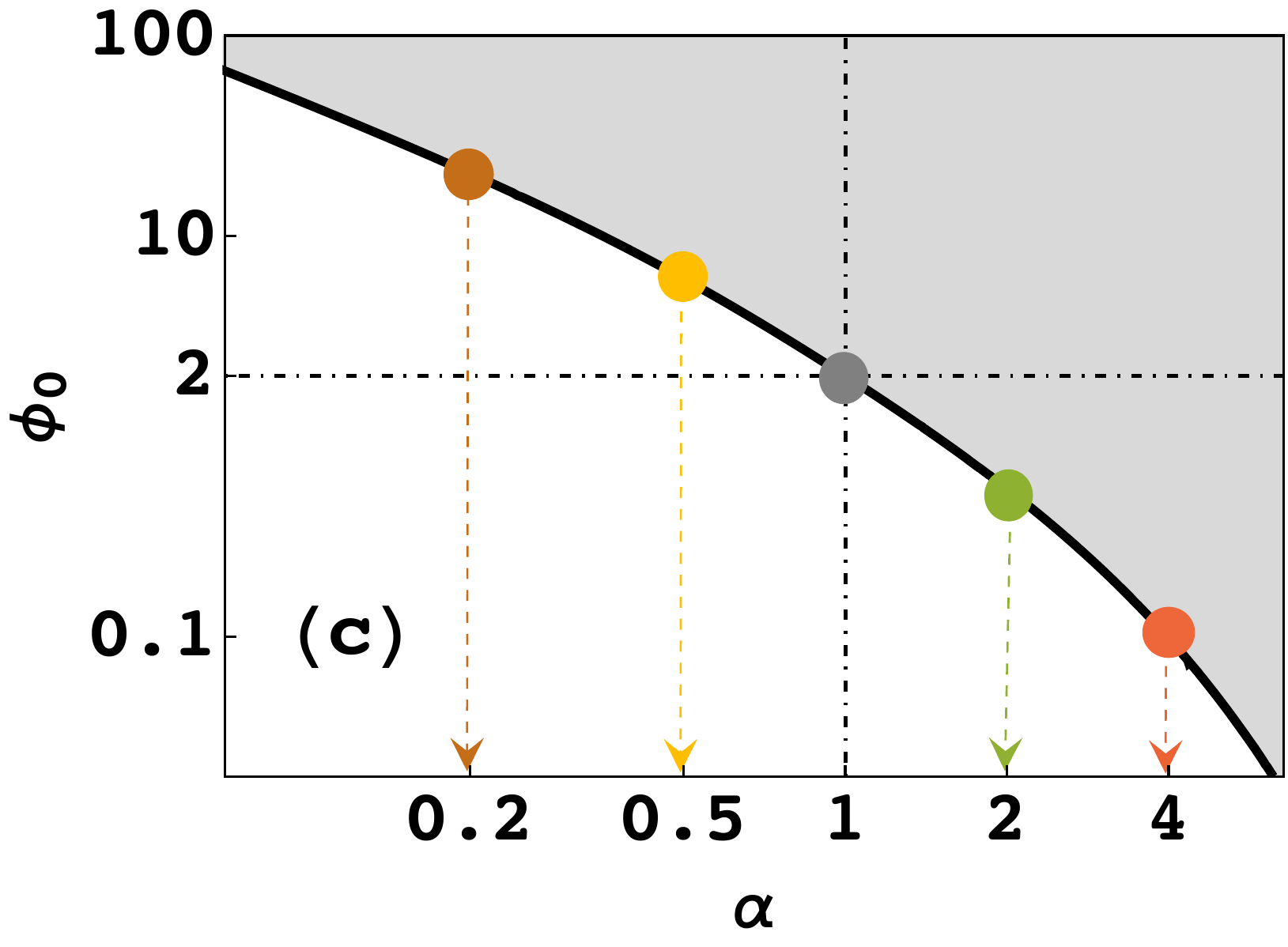}
\end{center}
\caption{Panel (a): A power-law potential $U(x)=U_0|x|^{\alpha}$ for different values of $\alpha$, and $U_0=1$. Panel (b): Graphical solution of \eref{eq:res_trans_plp} by plotting the ratio $\tau_2/2\tau_1^2$ (from the left hand side of \eref{eq:res_trans_plp}) vs. $\phi_0$ for different values of $\alpha$. The solutions, $\phi_0^{\star}$ (abscissa corresponding to the colored circles) correspond to the points where the resetting transition occurs. Panel (c): A phase diagram that shows the effects of resetting on the first-passage to the origin. The white region indicates the parameter space where the introduction of resetting accelerates first-passage. The shaded region displays the parameter space where resetting delays the first-passage. The points of the resetting transition are presented by plotting the numerical solutions of \eref{eq:res_trans_plp}) as a function of $\alpha$ (black line). Circles of the same color display the same values of $\alpha$ as in panel (b) . }
\label{FigPLP}
\end{figure*}
%%%%%%%%%%%%%%%%%%%%%%%%%%%%%%%%%%%%%%%%%%%%%
\indent
The logarithmic potential serves as an ideal example to illustrate the present framework, since in this case we can obtain an exact analytical expression of the critical value of $\nu$, which governs the resetting transition. For most of the commonly studied potentials, however, this is not the case. In what follows, we will discuss a broad class of potentials that increase monotonically with $x$. In general, for such a choice, the integrals given in \eref{eq:res_trans} can not be evaluated analytically and one needs to utilize numerical solutions instead. And yet, we show that here too an interplay between the thermal and the potential energy governs the resetting transition.\\
%%%%%%%%%%%%%%%%%%%%%%%%%%%%%%%%%%%%%%%%%%%%%
%\indent \underline{\it Monotonically increasing potentials:}\hspace{0.1cm} 
%%%%%%%%%%%%%%%%%%%%%%%%%%%%%%%%
\indent Consider a potential $U(x)=U_0f(x)$, where $U_0$ is the strength of the potential as before and $f(x)$ is a monotonically increasing function within the interval $x_a\le x<\infty$. Taking $x_a=0$, i.e, placing the absorbing boundary at the origin to simplify the analysis, we perform a general transformation of variable as $\phi\equiv\phi(x)\coloneqq\beta U_0f(x)$, so that $\phi^{\prime}\equiv \frac{d\phi }{dx}=\beta U_0 \frac{df(x)}{dx}$. \eref{eq:res_trans} can then be written as
\begin{widetext}
\begin{align}
\int_{\phi_a}^{\phi_0}d\phi(w)~\frac{e^{\phi(w)}}{\phi^{\prime}(w)}
\int_{\phi(w)}^{\phi_{\infty}}d\phi(v)~\frac{e^{-\phi(v)}}{\phi^{\prime}(v)}\int_{\phi_a}^{\phi(v)} d\phi(y)~\frac{e^{\phi(y)}}{\phi^{\prime}(y)}\int_{\phi(y)}^{\phi_{\infty}}d\phi(z)~\frac{e^{-\phi(z)}}{\phi^{\prime}(z)}
=\left[\int_{\phi_a}^{\phi_0}d\phi(y)~\frac{e^{\phi(y)}}{\phi^{\prime}(y)}\int_{\phi(y)}^{\phi_{\infty}}d\phi(z)~\frac{e^{-\phi(z)}}{\phi^{\prime}(z)}\right]^2,
\label{eq:res_trans1}
\end{align}
\end{widetext}
where $\phi_0\coloneqq\phi(x_0)\equiv\beta U_0f(x_0)$, $\phi_a\coloneqq\phi(x_a=0)\equiv\beta U_0f(x_a=0)$ and $\phi_{\infty}\coloneqq\phi(x\to \infty)\equiv\beta U_0f(x\to \infty)$. Note that in \eref{eq:res_trans1}, $\phi(i)$ for $i=w,v,y,z$ are integration variables, and the integrals depend only on the dimensionless parameters $\phi_0$, $\phi_a$ and $\phi_{\infty}$. Next, we utilize this to investigate the important case of power-law potentials, which further showcase the interplay between thermal and potential
energy in the resetting transition.\\
%%%%%%%%%%%%%%%%%%%%%%%%%%%%%%%%%%%%%%%%%%%%%%
%\indent\underline{\it Power-law potentials}: \hspace{0.1cm} 
%%%%%%%%%%%%%%%%%%%%%%%%%%%%%%%%%%%%%%%%%%%%%
\indent Consider a set of power-law potentials, $U(x)=U_0f(x)$ with $f(x)=|x|^{\alpha}$ and $\alpha>0$ [\fref{FigPLP}(a)]. For such potentials, using the notations $\phi_i\equiv\phi(i)$ for $i=w,v,y$ and $z$ we simplify \eref{eq:res_trans1} to obtain [see Supplemental Material for details]
\noindent
\begin{eqnarray}
\frac{\tau_2}{2\tau_1^2}&=&
\frac{\int_{0}^{\phi_0}d\phi_w\frac{{ e}^{+\phi_w}}{\phi_w^{1-\frac{1}{\alpha}}}\int_{\phi_w}^{\infty}d\phi_v\frac{{ e}^{-\phi_v}}{\phi_v^{1-\frac{1}{\alpha}}}\int_{0}^{\phi_v}d\phi_y\frac{{ e}^{+\phi_y}}{\phi_y^{1-\frac{1}{\alpha}}}\int_{\phi_y}^{\infty}d\phi_z\frac{{ e}^{-\phi_z}}{\phi_z^{1-\frac{1}{\alpha}}}}
{\left[\int_{0}^{\phi_0}d\phi_y\frac{{ e}^{+\phi_y}}{\phi_y^{1-\frac{1}{\alpha}}}\int_{\phi_y}^{\infty}d\phi_z\frac{{ e}^{-\phi_z}}{\phi_z^{1-\frac{1}{\alpha}}}\right]^2}\nonumber\\
&=&1,
\label{eq:res_trans_plp}
\end{eqnarray}
where $\phi_0=\beta U_0|x_0|^{\alpha}$ is the dimensionless potential energy of the particle at its initial position $x_0$.\\
\indent
For $\alpha=1$, evaluating the integrals in \eref{eq:res_trans_plp} we obtain
\begin{align}
\frac{\frac{1}{\phi_0^2}\left(1+\frac{2}{\phi_0}\right)}{\frac{2}{\phi_0^2}}
=1.
\label{eq:res_trans_linear2}
\end{align}
The non-trivial solution of \eref{eq:res_trans_linear2} is $\phi^{\star}_0=2$. Recalling the definition of the P\'eclet number, $Pe\coloneqq x_0(U_0\zeta^{-1})/2D=x_0\beta U_0/2$, we see that the condition $\phi^{\star}_0\coloneqq[\beta U_0 x_0]^{\star}=2$ is essentially the same as $Pe=1$, which agrees with previous work \cite{RayReuveniJPhysA}. \\
%%%%-----DWP-----%%%%%%%%%%%%%%%%%%%%%%%%
\begin{figure*}[ht!]
\begin{center}
\includegraphics[width=5.55cm]{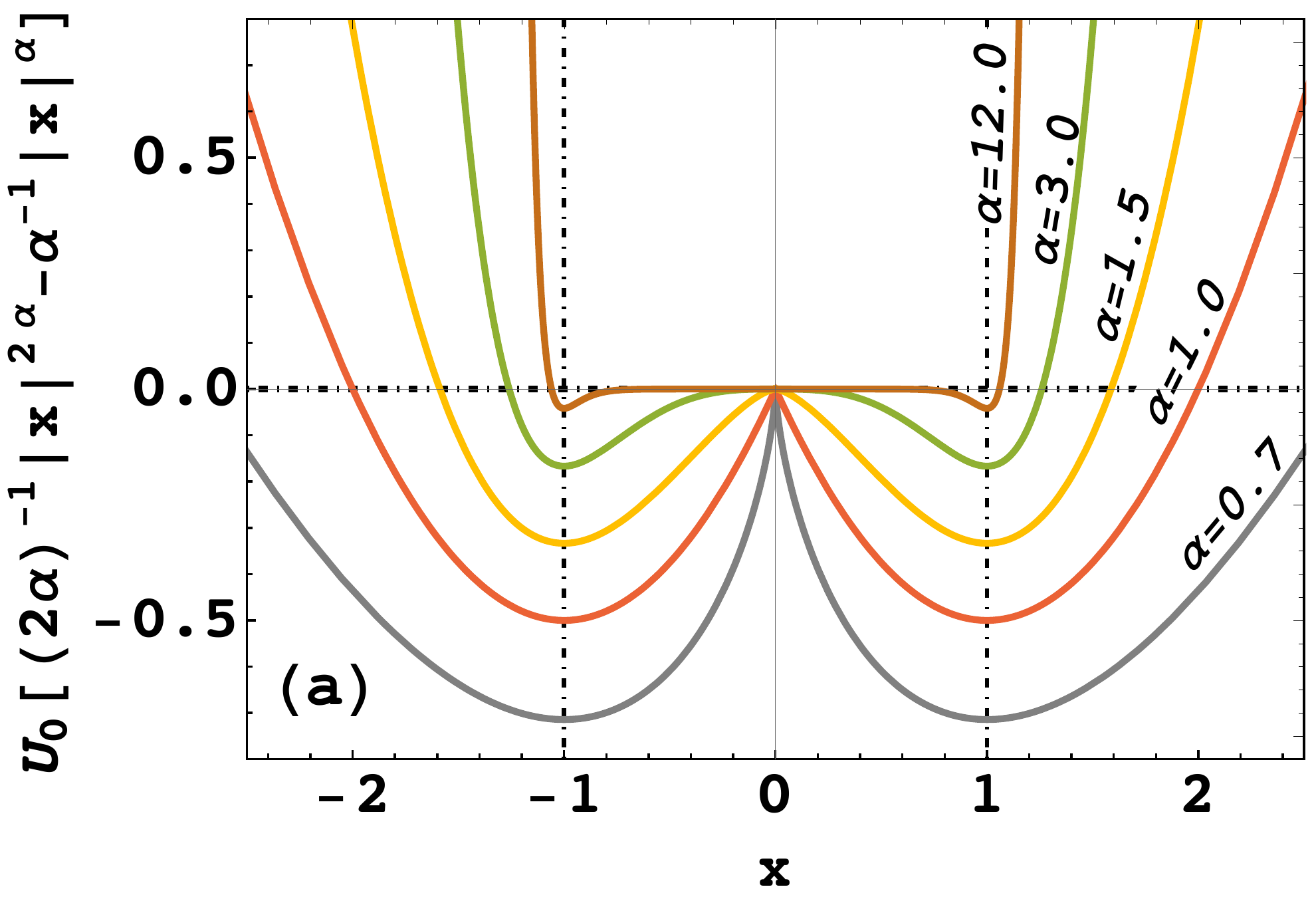}
\includegraphics[width=5.45cm]{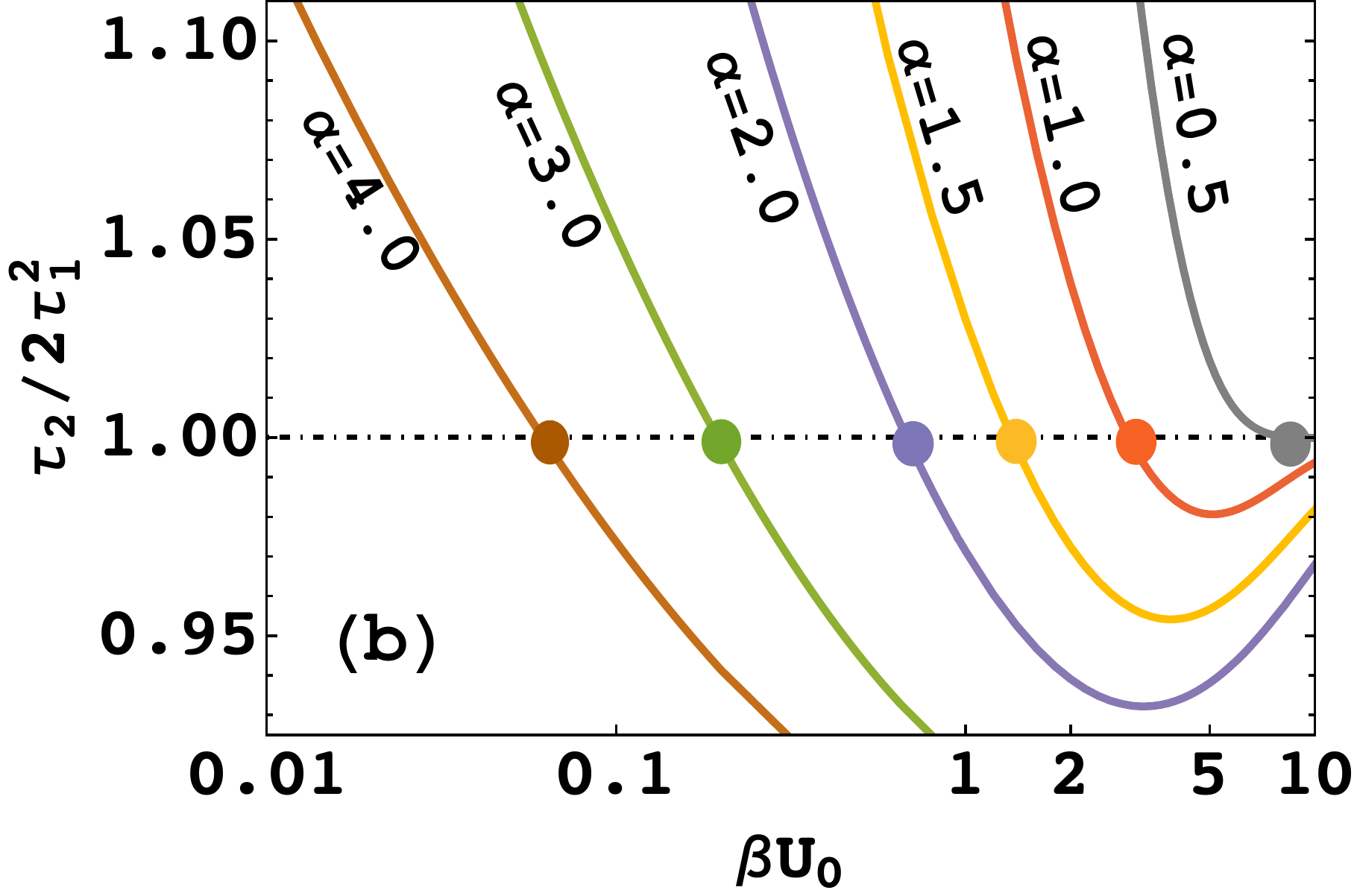}
\includegraphics[width=5.35cm]{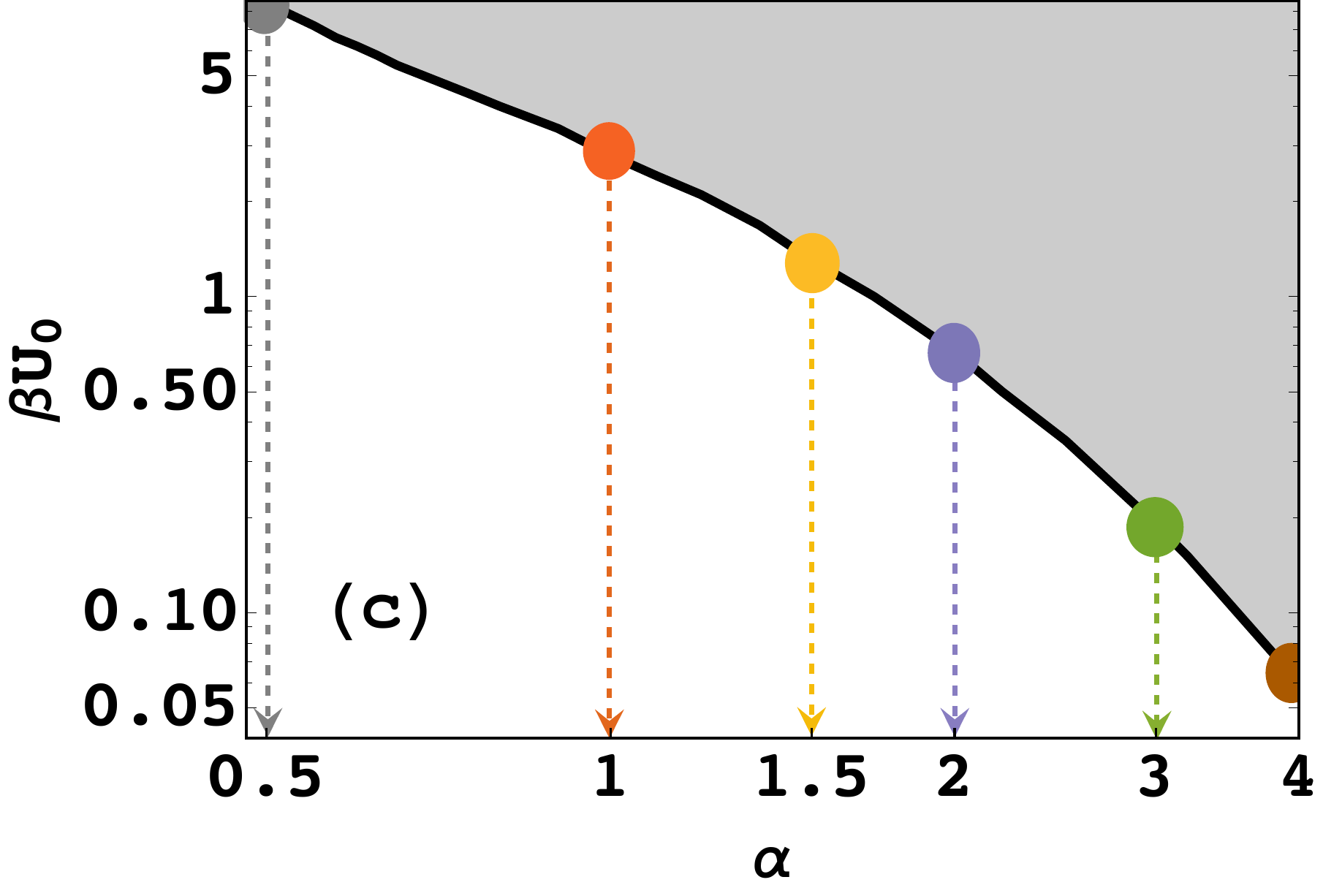}
\end{center}
\caption{Panel (a): Double-well potentials $U(x)=U_0\left[(2\alpha)^{-1}|x|^{2\alpha}-\alpha^{-1}|x|^{\alpha}\right]$ for $U_0=1$ and different values of $\alpha$. $U(x)$ owns minima at $x=\pm 1$ and a local maximum at $x=0$. Panel (b): Graphical solution of \eref{eq:res_trans_dwp} by plotting $\tau_2/2\tau_1^2$ from its left hand side vs. $\beta U_0$, for different values of $\alpha$. The solutions, $[\beta U_0]^{\star}$ (abscissa corresponding to the colored circles), denote the strength of the potential at the point of the resetting transition in units of the thermal energy. Panel (c): A phase diagram that exhibits the effects of resetting on barrier crossing. The white region indicates the parameter space where the introduction of resetting accelerates barrier crossing. The shaded region displays the parameter space where resetting delays the same. The points of the resetting transition are presented by the numerical solutions of \eref{eq:res_trans_dwp} as a function of $\alpha$ (black line). Circles of the same color display the same value of $\alpha$ as in panel (b). }
\label{FigNMP}
\end{figure*}
%%%%%%%%%%%%%%%%%%%%%%%%
\indent
For $\alpha\neq1$, the integrals in \eref{eq:res_trans_plp} can not be evaluated analytically. This, instead, we numerically evaluate for different values of $\alpha$ and plot the left hand side of \eref{eq:res_trans_plp} in \fref{FigPLP}(b), to present graphical solutions for this equation. The solutions, denoted $\phi^{\star}_0$, correspond to the resetting transition in each case. \fref{FigPLP}(c) displays a phase diagram that presents the effect of resetting on first-passage. For larger values of $\phi_0$ (shaded region), the potential strongly drives the particle towards the origin. Resetting interrupts such motion and thereby delays first-passage to the origin. In contrast, for smaller values of $\phi_0$ (white region), the potential is weak and the thermal energy dominates. Introduction of resetting thus accelerates first-passage to the origin. These two phases are separated by the points of the resetting transition, $\phi_0^{\star}$, which varies with $\alpha$ (black line).\\
\indent
Note that the points of the resetting transition above, i.e., $\tau_2=2\tau_1^2$, depend only on $\phi_0$, which is dimensionless by definition. Moreover, since $U(0)\coloneqq U_0f(0)=0$, we see that the transition depends solely on the energy released by the particle as it moves from it initial position $x_0>0$ to the absorbing boundary placed at the origin, in the units of the thermal energy $\beta^{-1}$. Comparing $\phi_0$ with the P\'eclet number \cite{redner,RayReuveniJPhysA} we see that $\phi_0$ can be considered as a generalized form of $Pe$ for a monotonically increasing (non-linear) potential of the form $U(x)=U_0f(x)$, which vanishes at the origin. Having thoroughly analysed this case, we now proceed to discuss the application of our framework to a set of non-monotonic potentials, viz., double-well potentials, as the final example of the present study. \\
%%%%%%%%%%%%%%%%%%%%%%%%%%%%%%%%%
%\underline{\it Double-well potentials:}
%%%%%%%%%%%%%%%%%%%%%%%%%%%%%%%%%
\indent
Consider a class of potentials   $U(x)=U_0\left(\frac{|x|^{2\alpha}}{2\alpha}-\frac{|x|^{\alpha}}{\alpha}\right)$, where $U_0$ denotes the strength of the potential as before and $\alpha>0$ [\fref{FigNMP}(a)]. For this particular choice, the potential $U(x)$ owns a couple of minima at $x=\pm 1$ irrespective of the value of $\alpha$, which is why one commonly addresses this set of potentials as {\it double-well} potentials. 
This class of non-monotonic potentials are frequently used in the context of rate theory and barrier crossing dynamics\cite{HanggiKramers,gammaitoni}. In what follows, we will explore the effect of resetting on barrier crossing by considering the first-passage of the diffusing particle from the minimum at $x_0=1$ to the top of the barrier at  $x_a=0$. \\
\indent
Recalling that $\phi(x)\coloneqq\beta U(x)\equiv\beta U_0\left(\frac{|x|^{2\alpha}}{2\alpha}-\frac{|x|^{\alpha}}{\alpha}\right)$ and utilizing \eref{eq:res_trans}, we see that the condition for resetting transition translates to
\begin{align}
\frac{\tau_2}{2\tau_1^2}=\frac{\int_{0}^{1}dw~e^{\phi(w)}
\int_{w}^{\infty}dv~e^{-\phi(v)}
\int_{0}^{v}dy~e^{\phi(y)}
\int_{y}^{\infty}dz~e^{-\phi(z)}}{\left[\int_{0}^{1}dy~e^{\phi(y)}
\int_{y}^{\infty}dz~e^{-\phi(z)}\right]^2}
=1.
\label{eq:res_trans_dwp}
\end{align}
The integrals in \eref{eq:res_trans_dwp} can not be evaluated analytically, however, it is not difficult to see that the results depend only on $\alpha$ and $\beta U_0$, the strength of the potential in terms of the thermal energy. We numerically calculate the ratio $\tau_2/2\tau_1^2$ from the left hand side of \eref{eq:res_trans_dwp} and plot it in \fref{FigNMP}(b) to graphically solve the equation for different values of $\alpha$. 
In \fref{FigNMP}(c), we construct a phase diagram by plotting the solutions, denoted $[\beta U_0]^{\star}$ vs. $\alpha$ (black line). The area under the curve (white region) shows the parameter space where resetting expedites barrier crossing (smaller values of $\beta U_0$), whereas the area above the curve (shaded region) marks the parameter space where resetting hinders the same (larger values of $\beta U_0$). Therefore, our analysis of barrier crossing in the double-well potential proves that even when the underlying system is not exactly solvable, the general framework outlined in the present paper can be utilized to locate the point of the resetting transition based on an interplay of two competing energy scales, the thermal and the potential energy.\\ 
%%%%%%%%%%%%%%%%%%%%%%%%%%%%%%%%%%%%%%%%%%%%%%%%
%\indent\underline{\it Concluding remarks:} \hspace{0.4cm}
%%%%%%%%%%%%%%%%%%%%%%%%%%%%%%%%%%%%%%%%%%%%%%%% 
\indent
In conclusion, we presented a general framework to determine the resetting transition point of diffusion in a potential. The framework applies to diffusion in arbitrary potentials, and it was illustrated with three different showcase potentials, viz., a regularized logarithmic potential, a set of power-law potentials of varying exponents and a set of double-well potentials of varying barrier heights, which are widely used to model a variety of first-passage processes. In most of the cases involving diffusion in a potential landscape with resetting,   solution of the Fokker Planck equation can be quite challenging. There, instead, our methodology can be adapted to identify the resetting transition point. More importantly, here we showed that the resetting transition can be understood in terms of an interplay between the thermal and the potential energies. These interpretations will strengthen the connection between theory and experiment in the field of stochastic resetting by paving new ways of analyzing experimental data and by generating new experimentally verifiable predictions.\\
%%%%%%%%%%%%%%%%%%%%%%%%%%%%%%%%%%%%%%%%%%%%%%%%
\noindent\underline{\it Acknowledgements:}\hspace{0.2cm}
%%%%%%%%%%%%%%%%%%%%%%%%%%%%%%%%%%%%%%%%%%%%%%%%
S. Ray acknowledges support from the Raymond and Beverly Sackler Center for Computational Molecular and Materials Science, Tel Aviv University, Israel and the DST-INSPIRE Faculty Grant, Govt. of India, executed at Indian Institute of Technology Tirupati (project no. CHY/2021/005/DSTX/SOMR). S. Reuveni acknowledges support from the Azrieli Foundation, from the Raymond and Beverly Sackler Center for Computational Molecular and Materials Science at Tel Aviv University, and from the Israel Science Foundation (grant No. 394/19).\\
\noindent\underline{\it Supporting Information Available:}\hspace{0.2cm} Details of the derivations are available in the Supporting Information at the end.

\vspace{2cm}

%%%%%%%%%%%%%%%%%%%%%%%%%%%%%%%%%%%%%%%%%%%%%%%%%%%%%%%%
\begin{widetext}
\begin{center}
\underline{\bf Supporting Information}
\end{center}
\renewcommand{\theequation}{S.\arabic{equation}}
\setcounter{equation}{0} 

%%%%%%%%%%%%%%%%%%%%%%%%%%%%%%%%%%%%%%%%%%%%%%%%%%%%%%%%%
\subsection{Derivation of the exact expressions of the first and second moments ($\tau_1$ and $\tau_2$) of the first-passage time from Eq.(3):}
%%%%%%%%%%%%%%%%%%%%%%%%%%%%%%%%%%%%%%%%%%%%%%%%%%%%%%%%%
\indent
Eq.(3) in the main text is a single variable, recursive, second-order differential equation in $\tau_n$, the nth moment of FPT. Utilizing the Einstein relation $D=(\beta\zeta)^{-1}$ in Eq.(3), where $\beta$ is the thermodynamic $\textit beta$, we get
\begin{align}
\dfrac{d^2 \tau_n}{d x_0^{2}}-\beta U^{\prime}(x_0)\dfrac{d \tau_n}{d x_0}+\left[\frac{n}{D}\right]\tau_{n-1}=0,
\label{eq:stnde}
\end{align}
\noindent
Setting $n=1$ in \eref{eq:stnde} and taking into account the fact that $\tau_0\coloneqq-\int_0^{\infty}dt\frac{\partial Q(t|x_0)}{\partial t}\equiv \left.Q(t|x_0)\right|_{t\to0}-\left.Q(t|x_0)\right|_{t\to\infty}=1$, we see that the resulting second order differential equation in $\tau_1$ can be reduced to a first order differential equation in $\tau_1^{\prime}\equiv d \tau_1/d x_0$ that reads
\begin{align}
\dfrac{d \tau_1^{\prime}}{d x_0}-\beta U^{\prime}(x_0)\tau_1^{\prime}+\frac{1}{D}=0.
\label{eq:st1de}
\end{align}
\noindent
With an integrating factor $e^{-\int dx_0\beta U^{\prime}(x_0)}\equiv e^{-\beta U(x_0)}$, \eref{eq:st1de} leads to the general solution
\begin{align}
\tau_1^{\prime}=\frac{1}{D}e^{+\beta U(x_0)} \left[C-\int_{x_a}^{x_0}dz~e^{-\beta U(z)}\right],
\label{eq:stp1_gsol}
\end{align}
\noindent
where $C$ is the arbitrary integration constant. Putting the %reflecting 
boundary condition $\lim_{x_0\to\infty} e^{-\beta U(x_0)}\tau_1^{\prime}(x_0)=0$ in \eref{eq:stp1_gsol}, we find $C=\int_{x_a}^{\infty}dz~e^{-\beta U(z)}$, and that in turn leads to the specific solution of \eref{eq:st1de}
\begin{align}
\tau_1^{\prime}=\frac{1}{D}e^{+\beta U(x_0)}\int_{x_0}^{\infty}dz~e^{-\beta U(z)}.
\label{eq:stp1_sol}
\end{align}
\noindent
Recalling that $x_a<x_0$, we now integrate \eref{eq:stp1_sol} within the interval $[x_a,x_0]$ and utilize the boundary condition $\tau_1({x_a})=0$ to obtain\cite{gardiner}
\begin{align}
\tau_1=\frac{1}{D}\int_{x_a}^{x_0}dy~e^{+\beta U(y)}\int_{y}^{\infty}dz~e^{-\beta U(z)}.
\label{eq:st1_sol}
\end{align}
\noindent
\eref{eq:st1_sol} presents an explicit expression for the mean FPT of a particle that diffuses in a potential $U(x)$ from its initial position $x_0$ to the absorbing boundary at $x_a$.\\
\indent
In a similar spirit, we can obtain an exact expression for $\tau_2$. For this, we set $n=2$ in \eref{eq:stnde} to obtain a second-order differential equation in $\tau_2$
\begin{eqnarray}
\dfrac{d^2 \tau_2}{d x_0^{2}}-\beta U^{\prime}(x_0)\dfrac{d \tau_2}{d x_0}=-\frac{2}{D^2}\left[\int_{x_a}^{x_0}dy~e^{+\beta U(y)}\int_{y}^{\infty}dz~e^{-\beta U(z)}\right],
\label{eq:st2de}
\end{eqnarray}
\noindent
where we utilized \eref{eq:st1_sol} to substitute for $\tau_1$. Solving for $\tau_2$ in the exact manner as before, we get
\begin{eqnarray}
\tau_2=\frac{2}{D^2}\int_{x_a}^{x_0}dw~e^{+\beta U(w)}
\int_{w}^{\infty}dv~e^{-\beta U(v)}
\int_{x_a}^{w}dy~e^{+\beta U(y)}\int_{y}^{\infty}dz~e^{-\beta U(z)}.
\label{eq:st2_sol}
\end{eqnarray}
\noindent
\eref{eq:st2_sol} presents an explicit expression for the second moment of the FPT.\\ \\
%%%%%%%%%%%%%%%%%%%%%%%%%%%%%%%%%%%%%%%%%%%%%%%%%%%%%%%%%%%%%%%%%%%%%%%%%%%%%%%%%%%%%%%%
\subsection{Transformation of variable $x\to \phi(x)\equiv \beta U(x)$ for the power-law potential $U(x)=U_0|x|^{\alpha}$: Derivation of Eq. (8):}
%%%%%%%%%%%%%%%%%%%%%%%%%%%%%%%%%%%%%%%%%%%%%%%%%%%%%%%%%%%%%%%%%%%%%%%%%%%%%%%%%%%%%%%%
As discussed in the main text before Eq. (7), we consider a general transformation of variable $x\to\phi(x)\coloneqq\beta U(x)$, which leads to $\phi(x)=\beta U_0|x|^{\alpha}$ for the power-law potential. Therefore, for $0\le x<\infty$ we get
\begin{align}
x\equiv\left[\frac{\phi(x)}{\beta U_0}\right]^{\frac{1}{\alpha}}.
\label{eq:svbt}
\end{align}
\noindent
Differentiating \eref{eq:svbt} with respect to $\phi(x)$
thus gives
\begin{align}
dx\equiv\frac{1}{\alpha(\beta U_0)^{\frac{1}{\alpha}}}\left[\frac{d\phi(x)}{\phi(x)^{1-\frac{1}{\alpha}}}\right].
\label{eq:sdvbt}
\end{align}
\noindent
Setting $x_a=0$, we can write \eref{eq:st1_sol} in terms of the transformed variable as
\begin{align}
\tau_1=\frac{1}{D\alpha^2(\beta U_0)^{\frac{2}{\alpha}}}\int_{0}^{\phi(x_0)}d\phi(y)~\frac{e^{+\phi(y)}}{\phi(y)^{1-\frac{1}{\alpha}}}\int_{\phi(y)}^{\infty}d\phi(z)~\frac{e^{-\phi(z)}}{\phi(z)^{1-\frac{1}{\alpha}}}.
\label{eq:st1_t}
\end{align}
\noindent
Note that $\phi(y)$ and $\phi(z)$ are integration variables in \eref{eq:st1_t}, and hence the value of the above double integral depends only on $\phi_0\equiv \phi(x_0)=\beta U_0|x_0|^{\alpha}$. To simplify notation, we use $\phi_z=\phi(z)$ and $\phi_y=\phi(y)$ hereafter. In a similar spirit as in the case of the mean FPT, \eref{eq:st2_sol} in terms of the transformed variable reads
\begin{align}
\tau_2=\frac{2}{D^2\alpha^4(\beta U_0)^{\frac{4}{\alpha}}}\int_{0}^{\phi_0}d\phi_w\frac{{ e}^{+\phi_w}}{\phi_w^{1-\frac{1}{\alpha}}}\int_{\phi_w}^{\infty}d\phi_v\frac{{ e}^{-\phi_v}}{\phi_v^{1-\frac{1}{\alpha}}}\int_{0}^{\phi_v}d\phi_y\frac{{ e}^{+\phi_y}}{\phi_y^{1-\frac{1}{\alpha}}}\int_{\phi_y}^{\infty}d\phi_z\frac{{ e}^{-\phi_z}}{\phi_z^{1-\frac{1}{\alpha}}},
\label{eq:st2_t}
\end{align}
\noindent
where $\phi_w=\phi(w)$ and $\phi_v=\phi(v)$. From Eqs.~(\ref{eq:st1_t}) and (\ref{eq:st2_t}), the condition for resetting transition, $\tau_2/2\tau_1^2=1$, can be obtained, which is presented by Eq. (8) in the main text.

\end{widetext}
\end{document}